\documentclass[conference]{IEEEtran}
\IEEEoverridecommandlockouts
\usepackage{cite}
\usepackage{amsmath,amssymb,amsfonts}
\usepackage{algorithmic}
\usepackage{graphicx}
\usepackage{textcomp}
\usepackage{xcolor}
\def\BibTeX{{\rm B\kern-.05em{\sc i\kern-.025em b}\kern-.08em
    T\kern-.1667em\lower.7ex\hbox{E}\kern-.125emX}}

\usepackage{amsmath} 
\usepackage{amsfonts}
\usepackage{hyperref}
\usepackage{url}

\setkeys{Gin}{width=0.9\linewidth,keepaspectratio}

\newcommand{\eqr}[1]{Eq.\thinspace(#1)}
\newcommand{\pfrac}[2]{\frac{\partial #1}{\partial #2}}

\newcommand{\mvec}[1]{\mathbf{#1}}
\newcommand{\gvec}[1]{\boldsymbol{#1}}

\newcommand{\gke}{\texttt{Gkeyll}}

\begin{document}

\title{Alias-free, matrix-free, \emph{and} quadrature-free discontinuous Galerkin algorithms for (plasma) kinetic equations
\thanks{This work was partially supported by U.S. Department of Energy contract No. DE-AC02-09CH11466 for the Princeton Plasma Physics Laboratory (AH) and a NASA Earth and Space Science Fellowship (Grant No. 80NSSC17K0428) to JJ. Both authors have contributed equally to the research presented here.}
}

\author{
\IEEEauthorblockN{Ammar Hakim}
\IEEEauthorblockA{\textit{Princeton Plasma Physics Laboratory} \\
\textit{Princeton University}\\  
Princeton, NJ \\
ahakim@pppl.gov}
\and
\IEEEauthorblockN{James Juno}
\IEEEauthorblockA{\textit{Institute for Research in Electronics and Applied Physics} \\
\textit{University of Maryland}\\
College Park, MD \\
jjuno@terpmail.umd.edu}
}

\maketitle

\begin{abstract}
Understanding fundamental kinetic processes is important for many problems, from plasma physics to gas dynamics. A first-principles approach to these problems requires a statistical description via the Boltzmann equation, coupled to appropriate field equations. In this paper we present a novel version of the discontinuous Galerkin (DG) algorithm to solve such kinetic equations. Unlike Monte-Carlo methods, we use a continuum scheme in which we directly discretize the 6D phase-space using discontinuous basis functions. Our DG scheme eliminates counting noise and aliasing errors that would otherwise contaminate the delicate field-particle interactions. We use modal basis functions with reduced degrees of freedom to improve efficiency while retaining a high formal order of convergence. Our implementation incorporates a number of software innovations: use of JIT compiled top-level language, automatically generated computational kernels and a sophisticated shared-memory MPI implementation to handle velocity space parallelization.
\end{abstract}


\section{Introduction}

Understanding fundamental kinetic processes is important in many
physical problems, from the astrophysics of self-gravitating systems,
to plasma physics and gas dynamics. Several recent satellite missions
observe the detailed structure of these systems, for example, the GAIA\cite{gaiamission}
mission that aims to collect the position, velocity and other data on
billions of stars in our galaxy, or the Parker Solar Probe\cite{parkerprobe} mission that is studying the detailed structure of the hot solar wind plasma that
permeates the solar system. Each of these missions aims to measure the \emph{phase-space} of the ``particles,'' e.g, stars in the
case of GAIA and electrons and ions in case of Parker Solar
Probe. The quality of data is unprecedented and promises to greatly
enrich our understanding. Clearly, large-scale simulation capability
is needed to interpret and understand the detailed physics revealed by
these measurements.

A near first-principles approach is to look at the statistical
description via the Boltzmann equation coupled to appropriate field
equations: Poisson equations for self-gravitating system and Maxwell's
equations for plasmas. The challenge in solving such systems is the
inherent nonlinearity due to the coupling of the particles and
fields, and that the particle dynamics evolves in 6D phase-space
(position-velocity), requiring a very careful treatment of all
field-particle interaction terms.

The fundamental object in the Boltzmann description is the
\emph{particle distribution function} $f(\mvec{z})$ that evolves in
phase-space $\mvec{z}\equiv (\mvec{x},\mvec{v})$. The particle distribution function is
defined such that $f(\mvec{x},\mvec{v})d\mvec{v}d\mvec{x}$ is the
number of particles in phase-space volume
$d\mvec{z} = d\mvec{v}d\mvec{x}$ at position-velocity location
$(\mvec{x},\mvec{v})$. The motion of particles comes about from
free-streaming and particle acceleration and is described by the
Boltzmann equation
\begin{align}
  \pfrac{f}{t} + \nabla_{\mvec{x}}\cdot(\mvec{v}f) +
  \nabla_{\mvec{v}}\cdot(\mvec{a}f) = C[f], \label{eq:Boltzmann}
\end{align}
where $\nabla_{\mvec{x}}$ and $\nabla_{\mvec{v}}$ are gradient
operators in configuration and velocity space respectively, and
$\mvec{a}$ is the acceleration. To treat the phase-space as a whole we
will often use
$\nabla_{\mvec{z}} \equiv (\nabla_{\mvec{x}},\nabla_{\mvec{v}})$ and
denote the phase-space flux as
$\gvec{\alpha} \equiv (\mvec{v},\mvec{a})$. The right-hand side of \eqr{\ref{eq:Boltzmann}} represents
collision terms that redistribute the particles in velocity space, but
in a manner that conserves density, momentum and energy. Even though the streaming of particles, $ \nabla_{\mvec{x}}\cdot(\mvec{v}f)$, in \eqr{\ref{eq:Boltzmann}} is linear, the collisions and coupling to the fields
via the acceleration, determined by velocity moments of the distribution
function, makes the complete particle$+$field equations a highly
nonlinear, integro-differential, 6D system.

The high dimensionality of \eqr{\ref{eq:Boltzmann}} means that for most problems,
especially in 6D, one requires the largest computational resources
one can muster. In this paper we present a novel version of the
discontinuous Galerkin (DG) algorithm to solve such kinetic
equations. Unlike traditional and widely-used Monte Carlo methods, such as the particle-in-cell (PIC)
method for plasmas, we use a \emph{continuum scheme} in which we directly
discretize the 6D phase-space using discontinuous basis functions. A
continuum scheme has the advantage that the counting noise inherent in
PIC methods is eliminated, however, at higher computational
complexity. Once the basis set and a numerical flux function are
determined, we compute all volume and surface terms in the DG
algorithm \emph{exactly}, eliminating all aliasing errors that would
otherwise contaminate the delicate field-particle interactions. This elimination of aliasing errors
is a critical aspect of capturing the physics, both in the linear and
nonlinear regimes.

We use modal basis functions (of the Serendipity
family\cite{Arnold:2011}) with reduced degrees of freedom (DOF) to
improve efficiency while retaining a high formal order of
convergence. Further, use of a computer algebra system (CAS) allows us
to compute all integrals analytically, and orthonormalization of the
basis leads to very sparse update kernels minimizing FLOPs and
eliminating all tensor-tensor products and explicit quadratures.

We extend previous work\cite{Juno:2018}, where the authors presented a \emph{nodal}
DG algorithm to solve the Boltzmann equation in the context of plasma
physics.
In the plasma physics context, the Boltzmann equation, coupled to Maxwell's equations, forms the Vlasov-Maxwell system of equations, in which charged particles evolve in
self-consistent electromagnetic fields. For the Vlasov-Maxwell system of equations, the acceleration vector
is given by $\mvec{a} = q(\mvec{E}+\mvec{v}\times\mvec{B})/m$, where
$q$ and $m$ are particle charge and mass, and $\mvec{E}$ and
$\mvec{B}$ are electric and magnetic fields, determined from Maxwell's
equations. The particle contribution to the electromagnetic fields is determined from the 
plasma currents that appears in Ampere's law. The work of \cite{Juno:2018} showed that a 
DG scheme can conserve the mass and, when using central fluxes for
Maxwell equations, total energy (particle$+$field) exactly. Importantly though, unlike the
case of fluid problems (Euler, Navier-Stokes, or magnetohydrodynamics equations), there
is no explicit energy equation that is evolved. In fact, the energy
(as discussed in Section II) depends on \emph{moments} of the
distribution function as well as the $L_2$-norm of the electromagnetic
field. Hence, ensuring both the accuracy of the evolution of the energy, and that the energy is conserved, is not trivial and care is
needed to maintain energy conservation. The \emph{modal} DG scheme presented
here does not change the properties proved in\cite{Juno:2018}, but it does greatly
improve the efficiency and scalability of the DG algorithm, while
maintaining all the scheme's favorable properties.

Our algorithms are implemented in the open-source \gke\cite{gkeyllgit,gkeylldocs} code that
incorporates a number of software innovations: use of JIT compiled
top-level language, CAS generated computational kernels, and a
sophisticated shared-memory MPI implementation to handle velocity
space parallelization. We have obtained sub-quadratic scaling of the
computational kernels with DOFs per-cell and also good parallel
weak-scaling of the code on the Theta
supercomputer.

The modal, alias-free,
matrix-free, and quadrature-free DG algorithm presented here has also been applied to the
discretization of Fokker-Planck equations \cite{Hakim:2020}.
We note though that, to our knowledge, this paper describes the first instance of the application of a modal DG algorithm which is simultaneously alias-free,
matrix-free, and quadrature-free to kinetic equations, especially nonlinear kinetic equations.
In the rest of the paper we describe some aspects of our schemes and
innovation we have made to make high-dimensional problems within
reach, at least on large supercomputers.

\section{Modal discontinuous Galerkin algorithm}
As context for the fundamental algorithmic advancement of this paper, we briefly review the ingredients of a discontinuous Galerkin scheme.
To construct a DG discretization of a partial differential equation (PDE) such as the kinetic equation, we discretize our phase space domain into grid cells, $K_j$, multiply the kinetic equation by test functions $w$, and integrate the phase space gradient by parts to construct the \emph{discrete-weak form},
\begin{align}
  \int_{K_j} w\pfrac{f_h}{t} \thinspace d\mvec{z} + 
  \oint_{\partial K_j} & w^- \mvec{n}\cdot\hat{\mvec{F}}  \thinspace dS \notag \\
  & - \int_{K_j} \nabla_{\mvec{z}} w \cdot \gvec{\alpha}_h f_h \thinspace d\mvec{z} = 0. 
\end{align}
The discrete-weak form is then evaluated in each grid cell $K_j$ and for every test function $w(\mvec{z})$ in a chosen basis expansion, with the discrete representation of the particle distribution defined as
\begin{align}
    f_h(\mvec{z}, t) = \sum_{i=1}^{N_p} f_i(t) w_i(\mvec{z}),
\end{align}
for the $N_p$ test functions which define our basis.
We likewise have a discrete representation for the phase space flux, $\gvec{\alpha}_h$, which, for example, looks like
\begin{align}
    \gvec{\alpha}_h = \left ( \mvec{v}, \frac{q}{m} [\mvec{E}_h + \mvec{v} \times \mvec{B}_h ] \right ),
\end{align}
for the particular Boltzmann equation for the evolution of a collisionless plasma, i.e., the Vlasov equation.
The numerical flux function, $\hat{\mvec{F}}$, is some suitably appropriate prescription for the interface fluxes, such as upwind fluxes, in analogy with traditional finite volume methods.
In contrast to finite volume methods though, the numerical flux function has its own basis expansion, e.g., for central fluxes,
\begin{align}
    \hat{\mvec{F}} = \frac{1}{2} \left (\gvec{\alpha}_h^+ f_h^+ + \gvec{\alpha}_h^- f_h^- \right), 
\end{align}
where the superscript $-(+)$ denote the basis expansions of $\gvec{\alpha}_h$ and $f_h$ evaluated just inside (outside) the cell interface.

While the discrete-weak form is a mathematically complete formulation of the DG algorithm, to translate the discrete-weak form into code, a suitable choice of basis functions for $w(\mvec{z})$ must be made to evaluate the integrals in the discrete-weak form.
Restricting ourselves to polynomial bases, a conventional approach in the application of DG methods to hyperbolic PDEs is a \emph{nodal} basis, wherein the basis set is defined by a set of polynomials whose values are known at nodes.
An example nodal basis in 1D is
\begin{align}
    f_h(x, t) = \sum_{k=1}^{N_p} f_k(\xi_k, t) \ell_k(x),
\end{align}
where $\ell_k$ are the Lagrange interpolating polynomials,
\begin{align}
     \ell_k(x) = \prod_{j=1, j\neq k}^{N_p} \frac{x - \xi_j}{\xi_k - \xi_j},
\end{align}
and $\xi_k$ are the $k$ nodes by which the polynomials are defined.
Because the polynomials are defined to take a value of one at one node and zero at all other nodes, the coefficients $f_k$ are thus known at the specific set of nodes.

Nodal bases are common in the DG literature because of the computational advantages they provide for many applications, most especially the simplification of many of the integrals if one substitutes products and other nonlinear combinations of basis functions as
\begin{align}
    \gvec{\alpha}_h(\mvec{z},t) f_h(\mvec{z},t) \approx \gvec{\alpha}_k (\gvec{\xi}_k, t) f_k (\gvec{\xi}_k, t).
\end{align}
Such a simplification reduces the number of operations required to evaluate the discrete-weak form and numerically integrate the PDE of interest, but at a cost: aliasing errors are introduced into the solution since nonlinear combinations of the nodal basis set are not contained in the basis \cite{Hesthaven:2007, Hindenlang:2012}.
These aliasing errors have been studied in the context of fluids equations, such as the Euler equations, the Navier-Stokes equations, or the equations of magnetohydrodynamics, where it is found that these aliasing errors can have a destabilizing effect \cite{Kirby:2003}.
However, the computational gains from the simplification of the integrals are large enough that significant effort has been spent on mitigating these errors with filtering and artificial dissipation \cite{Fischer:2001, Gassner:2013b, Flad:2016, Moura:2017} or split-form formulations\footnote{In the split-form formulation, conservative and non-conservative forms of the equation at the continuous level are averaged to produce a different (but mathematically equivalent), but ultimately more computationally favorable, equation to discretize.} \cite{Gassner:2013a, Gassner:2014, Gassner:2016a, Gassner:2016b, Flad:2017, Winters:2018}.
Because fluid equations involve explicit conservation relations and the aliasing errors manifest in the smallest scales and highest wavenumbers, there is far less concern that mitigation techniques such as filtering or artificial dissipation will lower the quality of the solution, at least at scales above the resolution of the simulation.
The ability to control aliasing errors while maintaining the favorable computational complexity of a nodal scheme is of tremendous utility for the simulation of large scale problems in computational fluid dynamics.

Unfortunately, these aliasing errors lead to uncontrolled numerical instabilities for kinetic equations.
To determine the source of these alias-driven numerical instabilities, we consider the example of the Vlasov equation for the evolution of a collisionless plasma. When employing at least piecewise quadratic polynomials, the DG discretization involves the evolution of the $|\mvec{v}|^2$ moment of the particle distribution function.
But the $1/2 \thinspace m|\mvec{v}|^2$ velocity moment is the particle energy, whose evolution is given by
\begin{align}
    & \frac{d}{dt} \left (\sum_j \int_{K_j} \frac{1}{2} m |\mvec{v}|^2 f_h \thinspace d\mvec{z} \right ) - \sum_j \int_{K_j} \nabla_{\mvec{z}} \left (|\mvec{v}|^2 \right ) \cdot \gvec{\alpha}_h f_h \thinspace d\mvec{z} \notag \\
    & = \frac{d}{dt} \left (\sum_j \int_{K_j} \frac{1}{2} m |\mvec{v}|^2 f_h \thinspace d\mvec{z} \right ) - \sum_j \int_{\Omega_j} \mvec{J}_h \cdot \mvec{E}_h \thinspace d\mvec{x}, \label{eq:discreteEnergyExchange}
\end{align}
where we have summed over all cells to eliminate the surface term, as in \cite{Juno:2018}, and substituted for the volume term the discrete exchange of energy between the particles and the electromagnetic fields, $\mvec{J}_h \cdot \mvec{E}_h$.
To obtain \eqr{\ref{eq:discreteEnergyExchange}}, we have integrated over velocity space and reduced the second integral to an integration over the configuration space cell $\Omega_j$.

In order for the substitution in \eqr{\ref{eq:discreteEnergyExchange}} to be valid, the integrations of the surface and volume terms must be performed exactly, or at least to a high precision, lest the aforementioned aliasing errors manifest themselves as the ``energy content'' of the velocity moments being transported in uncontrolled and undesirable ways.
It would be nigh impossible to correct the rearrangement of the ``energy content'' of the basis expansion in a physically reasonable way, much less a stable way, because these errors are entering at all scales and in both fields and particles, and destroying a fundamental property of the equation system: the exchange of energy between the plasma and electromagnetic fields is given by $\mvec{J}_h \cdot \mvec{E}_h$.
If we cannot safely apply standard techniques such as filtering to mitigate aliasing errors, we must then eliminate these errors in their entirety.

Eliminating aliasing errors with a nodal basis comes at a high cost though.
The use of numerical quadrature, even anisotropic quadrature as in \cite{Juno:2018}, leads to a computational complexity $\mathcal{O}(N_q N_p)$, where $N_q$ is the number of quadrature points required to exactly integrate the nonlinear term(s) in the kinetic equation. 
The number of quadrature points exponentially increases with dimensionality, leading to an incredibly expensive numerical method for five and six dimensional problems.

We can gain insight into how to manage this cost, while respecting our requirement to completely eliminate of aliasing errors, by considering the fundamental operation of our DG method.
Substitution of the full expansions for the phase space flux, $\gvec{\alpha}_h$, and distribution function, $f_h$, into the volume term in the discrete weak form gives us
\begin{align}
\int_{K_j} & \nabla_{\mvec{z}} w_l(\mvec{z}) \cdot \gvec{\alpha}_h(\mvec{z}, t) f_h(\mvec{z}, t) \thinspace d\mvec{z} = \\
& \sum_{m=1}^{N_p}\sum_{n=1}^{N_p}  \underbrace{\left (\int_{K_j} w_m(\mvec{z}) w_n(\mvec{z}) \nabla_{\mvec{z}} w_l(\mvec{z}) \thinspace d\mvec{z} \right )}_{\mathcal{C}_{lmn}} \cdot \gvec{\alpha}_m(t) f_n(t), \notag
\end{align}
where we have encompassed the spatial discretization in the evaluation and convolution of the entries in the tensor, $\mathcal{C}_{lmn}$.
If this tensor is dense, the convolution of $\mathcal{C}_{lmn}$ to evaluate the volume integral in the discrete-weak form will have a computational complexity of $\mathcal{O}(N_p^3)$, which would suffer the same curse of dimensionality as the use of numerical quadrature.

However, if $\mathcal{C}_{lmn}$ could be made sparse, this would correspond to a systematic reduction in the number of operations required to evaluate the volume integral, and thus reduce the number of operations to numerically integrate the kinetic equation with our DG method.
We can indeed sparsify $\mathcal{C}_{lmn}$ with the use of a \emph{modal, orthonormal} polynomial basis set, as many entries of the tensor will be zero if the basis functions $w(\mvec{z})$ are orthonormal.
In addition to the reduction in the number of operations required to evaluate the volume integral, the use of an orthonormal basis to sparsify $\mathcal{C}_{lmn}$ allows for a complete redesign of the algorithm to maximize performance on modern architectures.

\begin{figure*}
    \centering
    \includegraphics[width=0.9\textwidth]{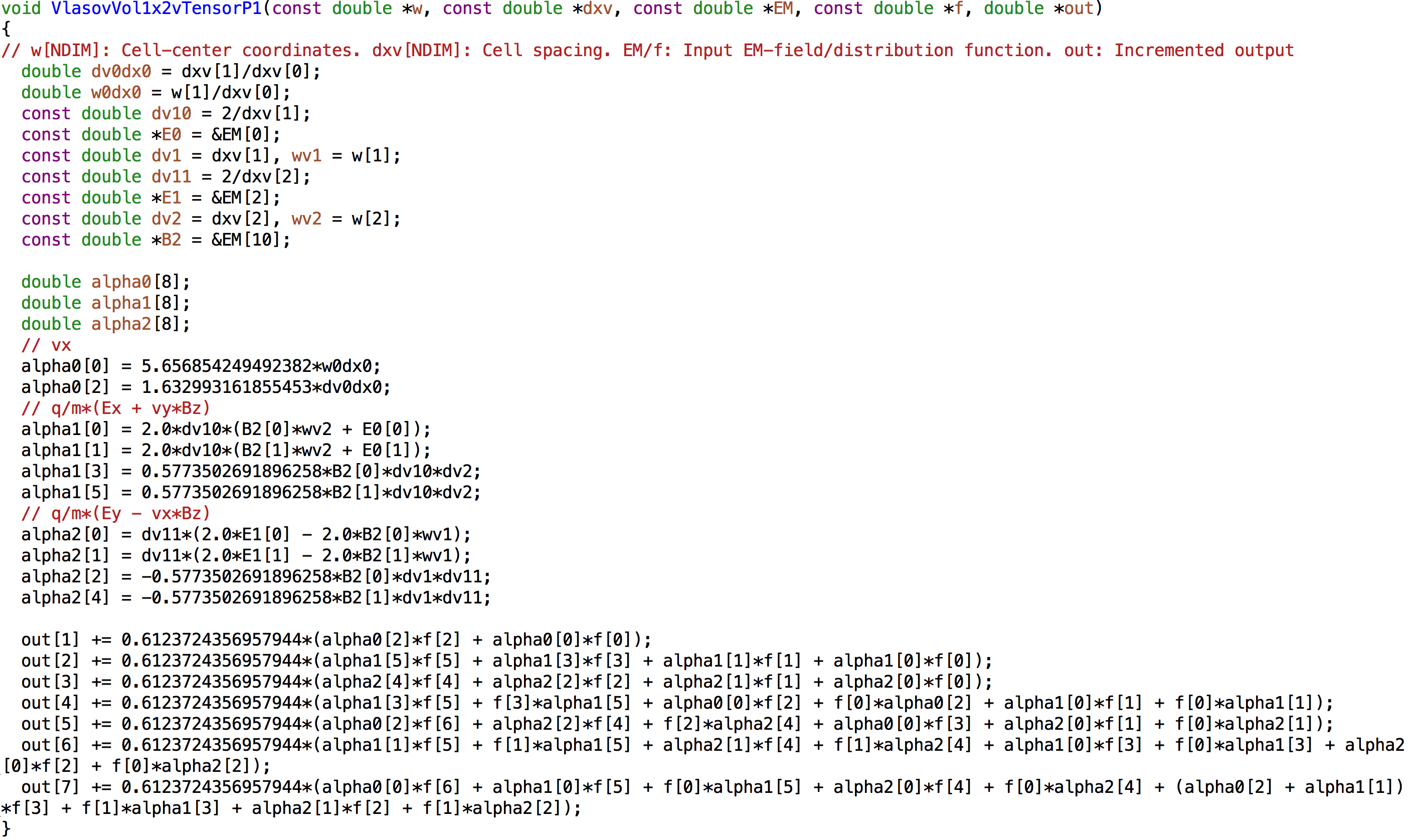}
    \caption{The computational kernel for the volume integral for the collisionless advection in phase space of the particle distribution function in one spatial dimension and two velocity dimensions (1X2V) for the piecewise linear tensor product basis. Note that this computational kernel takes the form of a C++ kernel that can be called repeatedly for each grid cell $K_j$ depending on the local cell center coordinate and the local grid spacing. Here, the local cell coordinate is the input ``const double w'' and the local grid spacing is the input ``const double dxv''. The out array is the increment to the right hand side due this volume integral contribution in a forward Euler time-step. To complete a forward Euler time-step for the evolution of the particle distribution function, for a given phase space cell, we require the surface contributions for the collisionless advection.}
    \label{fig:1X2V-p1-vol-update}
\end{figure*}

We now describe the principal algorithmic advancement of this paper: an alias-free, matrix-free, \emph{and} quadrature-free DG algorithm for kinetic equations.
By choosing a modal, orthonormal polynomial basis, we can symbolically integrate the individual terms in the tensor $\mathcal{C}_{lmn}$ and explicitly evaluate the sums which form the core of the update formulae. 
We construct computational kernels using the Maxima \cite{maxima} CAS to evaluate sums such as
\begin{align}
    \textrm{out}_l = \sum_{m=1}^{N_p}\sum_{n=1}^{N_p} \mathcal{C}_{lmn} \cdot \gvec{\alpha}_n f_m,
\end{align}
with similar computational kernels for the surface integrals.
We show an example computational kernel for the volume integral of the Vlasov equation in Figure~\ref{fig:1X2V-p1-vol-update} for the piecewise linear tensor product basis in one spatial and two velocity dimensions (1X2V).

Figure~\ref{fig:1X2V-p1-vol-update} shows a C++ computational kernel that can be called for every cell $K_j$ of a structured, Cartesian grid in phase space, as we are passing all the information required to the kernel to determine where we are physically in phase space, i.e., the local cell center coordinate and grid cell size.
The output of this computational kernel, the out array, forms a component of a system of ordinary differential equations,
\begin{align}
   \frac{d f_l}{dt} = \sum_{m=1}^{N_p} \mathcal{U}_{l m} \cdot \hat{\mvec{F}}_m (t) + \sum_{m=1}^{N_p}\sum_{n=1}^{N_p} \mathcal{C}_{l m n} \cdot \gvec{\alpha}_n (t) f_m (t),
\end{align}
where the operation $\mathcal{U}_{l m} \cdot \hat{\mvec{F}}_m (t)$ encodes the evaluation of the surface integrals on each surface of the cell and can also be pre-generated using a CAS\footnote{In the construction of this ordinary differential equation system, the matrix
\begin{align*}
    M_{kl} = \int_{K_j} w_k(\mvec{z}) w_l(\mvec{z}) d\mvec{z},
\end{align*}
must be inverted to solve for $df_l/dt$, but due to the choice of a modal, orthonormal basis this matrix is the identity matrix and thus requires no additional operations to invert.
}.
Given the computation of the surface and volume integrals in every cell, this system of ordinary differential equations can be discretized with an appropriate ODE integrator such as a strong-stability preserving Runge-Kutta (SSP-RK) method, as is done in \gke.
We note that we will likewise have computational kernels for Maxwell's equations, or another set of field equations such as Poisson's equations for self-gravitating systems, which must be evaluated at each stage of a SSP-RK method to complete the field-particle coupling.

Notably, the computational kernel in Figure~\ref{fig:1X2V-p1-vol-update} has no matrix data structure, much less the requirement to perform quadrature since we have already analytically evaluated the integrals which make up the entries of $\mathcal{C}_{lmn}$ with a CAS and written out the results to double precision. Further, we unroll all loops, eliminate common expressions and collect terms to ensure that the update uses fewer FLOPs\footnote{The problem of ensuring \emph{least} FLOP counts is difficult, and we apply most reasonably straightforward tricks we can think of. Certainly, a further reduction is likely possible and could be explored with more sophisticated optimization tools.}.
Using the local cell-center and grid spacing, we construct the phase space expansion of the phase space flux, $\gvec{\alpha}_h$, for each dimension, and then compute the convolution of the tensor $\mathcal{C}_{lmn}$ summed over each component of the phase space flux.
Thus, not only is the method alias-free because the integrals which form our spatial discretization have been evaluated to machine-precision, the method is also quadrature-free and matrix-free.
Such quadrature-free methods using orthogonal polynomials were studied in the early days of the DG method \cite{Atkins:1998,Lockard:1999} and are still applied to a variety of linear hyperbolic equations, such as the acoustic wave equation for studies of seismic activity, the level set equation, and Maxwell's equations \cite{Kaser:2006,Marchandise:2006,Koutschan:2012,Kapidani:2020}.
Even for alternative formulations of DG which do not seek to eliminate aliasing errors by exactly integrating the components of the discrete weak form, matrix-free implementations are desirable to reduce the memory footprint of the scheme \cite{Fehn:2019}.

To our knowledge, the construction of the modal, alias-free, matrix-free, and quadrature-free DG algorithm, henceforth abbreviated as simply the modal DG algorithm, is the first instance of such an algorithm design in the literature.
This particular algorithm design has numerous advantages, especially for nonlinear kinetic equations such as the Vlasov equation for collisionless plasma dynamics.
The sparseness of this modal DG algorithm leads to a reduction in the number of operations required to evaluate the volume and surface integrals, e.g., the computational kernel in Figure~\ref{fig:1X2V-p1-vol-update} has $\sim 70$ multiplications, whereas the update for numerical quadrature applied to an alias-free nodal basis has $\sim 250$ multiplications.
Critically, the modal DG algorithm both reduces the number of operations required to update the solution while respecting our requirement that we eliminate aliasing errors for stability and accuracy, thus providing an equally correct solution at a fraction of the cost.
In addition, the reduced memory footprint from requiring no matrix data structure and the unfolding of the tensor-tensor convolutions leads to additional performance improvements as the compiler can aggressively optimize the expressions.
To more precisely evaluate performance and determine quantitatively the computational complexity of the modal DG algorithm we will perform a numerical experiment in the next section.

\section{Computation complexity}\label{sec:compComplex}

\begin{figure*}
    \centering
    \includegraphics[width=0.47\textwidth]{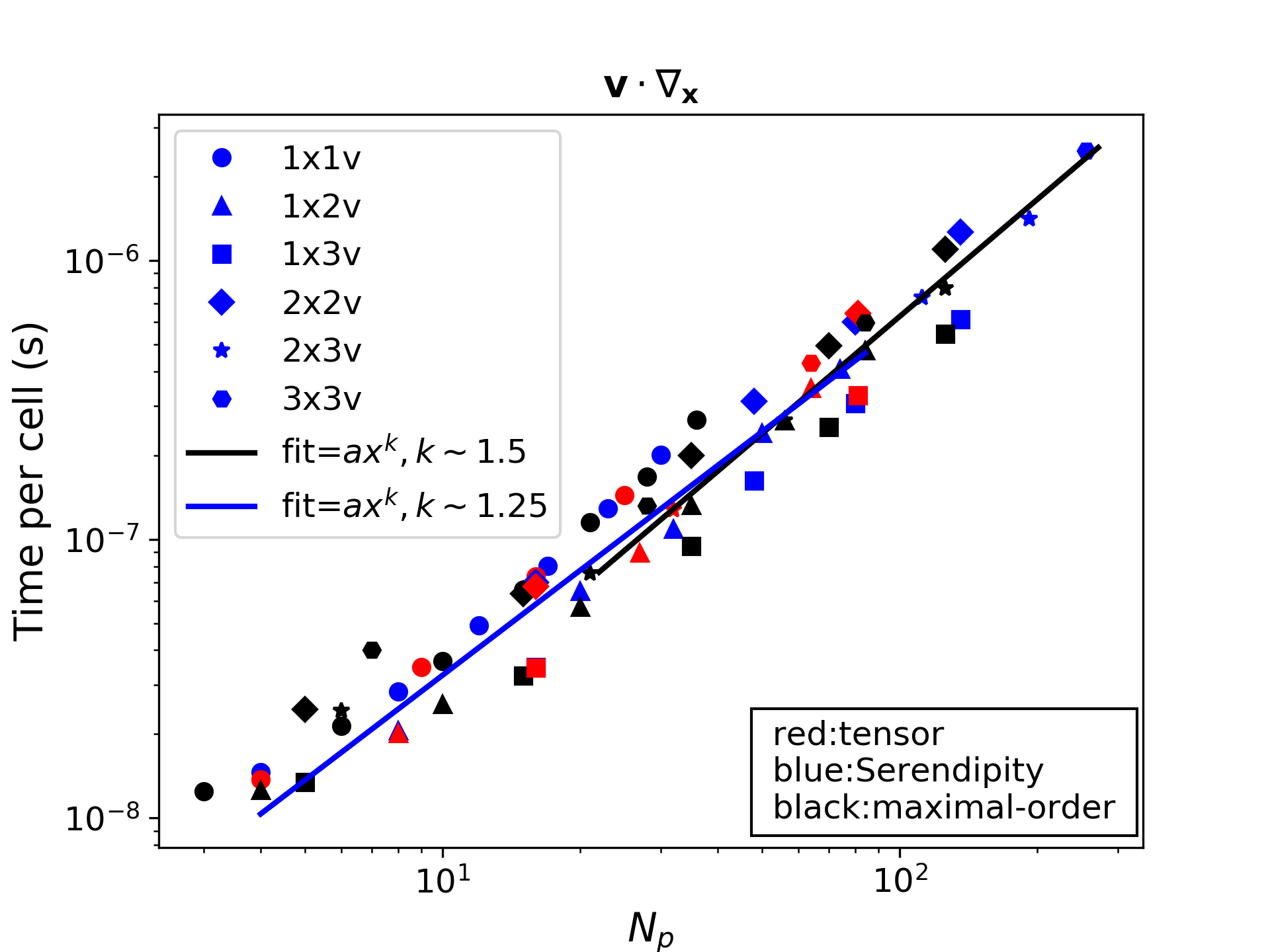}
    \includegraphics[width=0.47\textwidth]{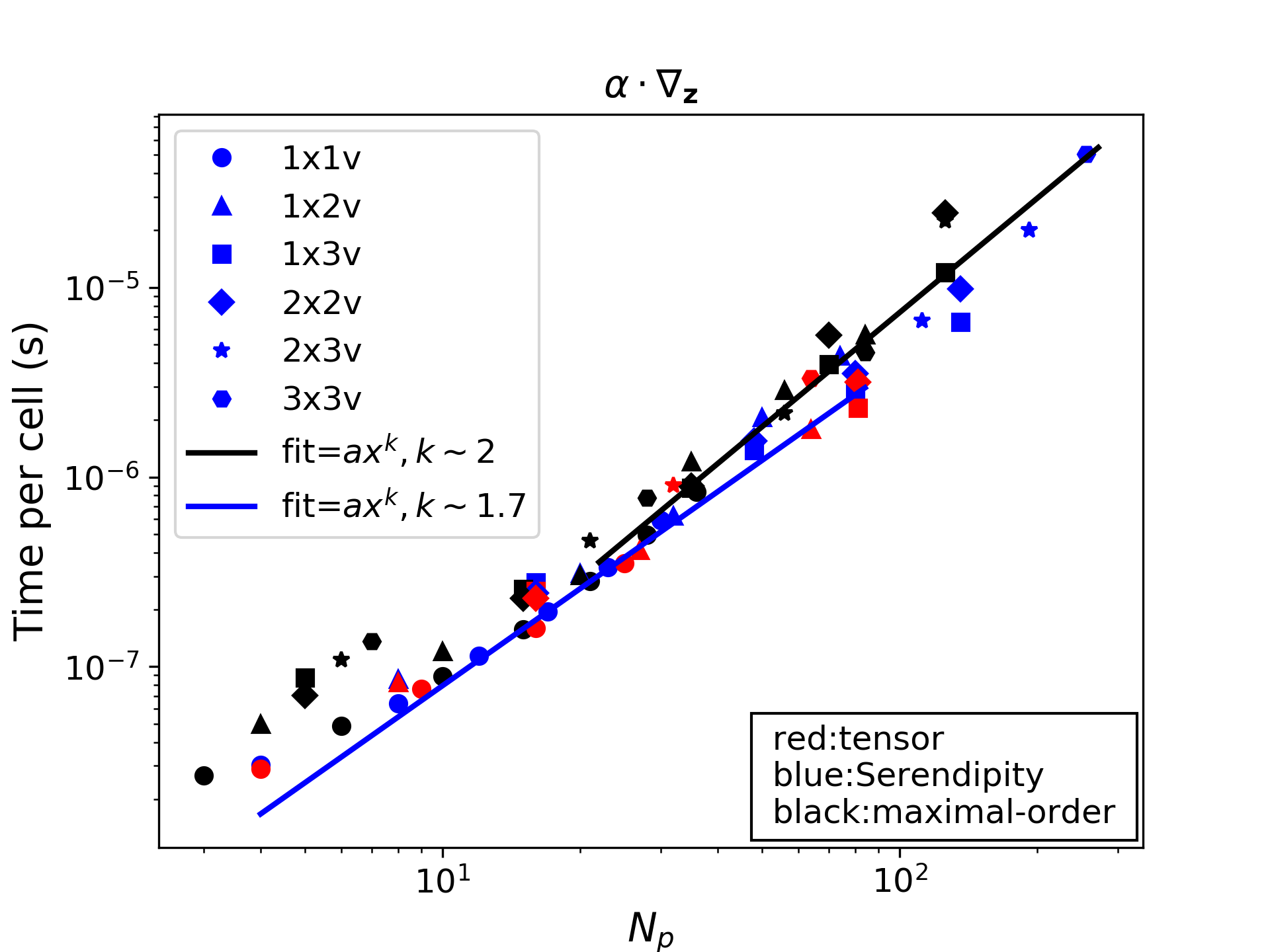}
    \caption{Scaling, i.e., the time to evaluate the update versus the number degrees of freedom, $N_p$, in a cell, of just the streaming term, $\gvec{\alpha} = (\mvec{v}, 0)$, (left) and the total, streaming and acceleration, update (right) for the Vlasov solver. The dimensionality of the solve is denoted by the relevant marker, e.g. 1x1v: one configuration space dimension and one velocity space dimension, and the three colors correspond to three different basis expansions: black: maximal-order, blue: Serendipity, and red: tensor. Importantly, this is the scaling of the \emph{full} update, for every dimension, i.e., the 3x3v, three configuration space and three velocity space dimensions, data points include the six dimensional volume integral and all twelve five dimensional surface integrals. In addition, we emphasize that the computational complexity is robust to the basis type. The cost of the method scales simply with the number of degrees of freedom within a cell, and no additional complexity is introduced when using the full tensor product basis versus the reduced Serendipity or maximal-order bases. \label{fig:algorithmscaling}}
\end{figure*}

Although we have evidence from the computational kernel presented in Figure~\ref{fig:1X2V-p1-vol-update} that the number of operations is indeed reduced compared to the use of numerical quadrature, we would like to determine generally how sparse the tensors required to update the discrete kinetic equation are.
We again take the example of the Vlasov equation, and in Figure~\ref{fig:algorithmscaling} show the results of a numerical experiment using the computational kernels generated from a variety of basis expansion and dimensionality combinations.
We show the time to evaluate the computational kernels in a phase space cell for just the streaming term, $\gvec{\alpha}_h = (\mvec{v}, 0)$ in the left plot of Figure~\ref{fig:algorithmscaling}, and the evaluation of the full phase space update, streaming and acceleration, in the right plot.
From the scaling of the cost to evaluate these computational kernels we can determine the computational complexity of the algorithm with respect to the number of degrees of freedom per cell, i.e., the number of basis functions in our expansion, $N_p$.

It is immediately apparent that even with the steepening of the scaling as the number of degrees of freedom increases there is at least some gain over the use of direct quadrature to evaluate the integrals in the discrete weak form because, at worst, the total, streaming plus acceleration, update scales roughly as $\mathcal{O}(N_p^2)$.
In fact, this scaling of, at worst $\mathcal{O}(N_p^2)$, is exactly the scaling obtained by under-integrating the nonlinear term in a nodal basis \cite{Hesthaven:2007,Hindenlang:2012}.
But critically, we have obtained this \emph{same} (or better) computational complexity while \emph{eliminating} aliasing errors from our scheme, as we require for stability and accuracy.

However, the improvement in the scaling is actually better than it first appears.
The scaling shown in Fig.~\ref{fig:algorithmscaling} is the cost scaling of the full update to perform a forward Euler step in a phase space cell, i.e., in six dimensions, three spatial and three velocity, the total update time in the right plot of Fig.~\ref{fig:algorithmscaling} is the time to compute the six dimensional volume integral plus the twelve required five dimensional surface integrals\footnote{Interestingly, the choice of orthonormal basis and analytically computing all integrals leads to the rather surprising result that the 6D volume integral is much cheaper than the surface integrals. In fact, the total cost of our algorithms is driven entirely by the surface integration costs.}. 
This means the scaling we are quoting is irrespective of the dimensionality of the problem, unlike in the case of the nodal basis, where the quadrature must be performed for every integral and there is a hidden dimensionality factor in the scaling.
In other words, in six dimensions, what at first may only seem like a factor of $N_q/N_p \sim 7$ improvement moving from a nodal to an orthonormal, modal representation is in fact a factor of $d N_q/N_p \sim 40$ improvement in the scaling once one includes the dimensionality factor, up to the constant of proportionality of the scaling. 
Of course, one must also compare the size of the constant of proportionality multiplying both scalings to accurately compare the reduction in the number of operations and improvement in the overall performance, since said constant of proportionality can either tell us the picture is much rosier, that in fact the improvement in performance is larger than we expected, or much more dire, that the improvement in the scaling is offset by a larger constant of proportionality.

To determine the constant of proportionality, we perform a more thorough numerical experiment and compare the cost of the alias-free nodal scheme in \cite{Juno:2018} and the alias-free modal scheme presented here for a complete collisionless Vlasov--Maxwell simulation.
Both schemes are implemented in \gke\, which will be discussed more extensively in Section~\ref{sec:Gkeyll}.
We consider the following test: a 2X3V computation done with both the nodal and the modal algorithms, with a detailed timing breakdown of the most important step of the algorithm, the Vlasov time step. 
The reader is referred Table \ref{table:VlasovSummaryTable} for a summary of the following two paragraphs if they wish to skip the details of the computer architecture and optimizations employed.
Both computations are performed in serial on a Macbook Pro with an \textbf{Intel Core i7-4850HQ (``Crystal Well'')} chip, the same architecture on which the scaling analysis in Figure~\ref{fig:algorithmscaling} was performed. 
The only optimization in the compilation of both algorithms is \textbf{``O3''} and both versions of the code are compiled with the C++ \textbf{Clang 9.1} compiler. 

Specific details of the computations are as follows: a $16^2 \times 16^3$ grid, with polynomial order two, and the Serendipity basis, 112 degrees of freedom per cell. 
The two simulations were run for a number of time-steps to allow us to more accurately compute the time per step of just the Vlasov solver, as well as the time per step of the complete simulation. 
The time-stepper of choice for this numerical experiment is the three-stage, third order, SSP-RK method \cite{Shu:2002,Durran:2010}.
To make the simulations as realistic as possible in terms of memory movement, we also evolve a ``proton'' and ``electron'' distribution function, i.e., we evolve the Vlasov-Maxwell system of equations for two plasma species.

To make the comparison as favorable as possible for the nodal algorithm, we also employ the highly tuned Eigen linear algebra library, \textbf{Eigen 3.3.4} \cite{eigen}, to perform the dense matrix-vector multiplies required to evaluate the higher order quadrature needed to eliminate aliasing errors in the nodal DG discretization. 
And we note that the nodal algorithm is optimized to use anisotropic quadrature (just high enough to eliminate aliasing) and uses only the surface basis functions in the surface integral evaluations, so we are minimizing the number of operations as much as possible to reduce the cost of the alias-free nodal scheme.

The results are as follows: 
for the \emph{nodal} basis, the computation required \textbf{1079.63} seconds per time step, of which \textbf{1033.89} seconds were spent solving the Vlasov equation. 
The remaining time is split between the computation of Maxwell's equations, the computation of the current from the first velocity moment of the distribution function to couple the particles and the fields, and the accumulation of each Runge-Kutta stage from our three stage Runge-Kutta method. 
For the \emph{modal} basis, the computation required \textbf{67.43} seconds per time step, of which \textbf{60.34} seconds were spent solving the Vlasov equation.

In the nodal case, we emphasize that we achieve a reasonable CPU efficiency, and the nodal timings are not a matter of poor implementation.
We estimate the number of multiplications in the alias-free nodal algorithm required to perform a full time-step is $\sim 3e12$, three trillion, once one considers the fact that we are evolving two distribution functions with a three-stage Runge--Kutta method.
One thousand seconds to perform three trillion multiplications corresponds to an efficiency of $\sim 3e9$ flops per second (3 GFlops/s).
This estimate is within 50 percent of the measured efficiencies of Eigen's matrix-vector multiplication routines for \textbf{Eigen 3.3.4} on a similar CPU architecture to the one employed for this test \cite{eigen}, so we argue that the cost of the alias-free nodal algorithm is due to the number of operations required and not an inefficient implementation of the algorithm\footnote{Note that this estimate for the efficiency of the nodal scheme, as well as the measured efficiency of \textbf{Eigen 3.3.4}'s matrix-vector multiplication routines, are for a serial computation on one core. Although the theoretical peak efficiency of an Intel Core i7-4850HQ is $\sim 150$ GFlops/s, we are not leveraging the four cores and eight hardware threads for the purposes of this cost analysis.
As we will show in Section~\ref{sec:Gkeyll}, both the nodal and modal algorithms achieve good parallel scaling since these methods require only local communication, and we could thus scale the time per step down and measured floating point efficiency up performing this cost analysis test in parallel.}.

Given these findings for the performance of the modal and nodal algorithms, it is then worth discussing how this improvement in the timings using the modal algorithm compares with our expectations.
From the scaling of the modal basis, we would anticipate the gain in efficiency in five dimensions would be around a factor of twenty, a factor of four from the reduction in the scaling from $\mathcal{O}(N_q N_p)$ to $\mathcal{O}(N_p^2)$, and an additional factor of five from the latter scaling containing all of the five dimensional volume integrals and the ten four dimensional surface integrals.
We can see that the gain in just the Vlasov solver is $\sim 17$, while the gain in the overall time per step is $\sim 16$, not quite as much as we would naively expect, but still a sizable increase in the speed of the Vlasov solver. 
The reduction in the overall time is due to the fact that, while the time to solve Maxwell's equations and compute the currents to couple the Vlasov equation and Maxwell's equations is reduced, these other two costs, in addition to the cost to accumulate each Runge-Kutta stage, are not sped-up as dramatically as the time to solve the Vlasov equation is.
\begin{table}
\begin{center}
\begin{tabular}{| l | l | l |}
\hline
\textbf{Computer} & \textbf{Architecture} & \textbf{Compiler} \\ \hline
MacBook Pro & Intel Core i7-4850HQ  & Clang 9.1 C++ \\
(High Sierra OS) & (``Crystal Well'') & \\ \hline
\textbf{Optimization Flags} & \textbf{Grid Size} & \textbf{Polynomial Order}  \\ \hline
``O3,'' & $16^2 \times 16^3$ & Serendipity quadratic, \\ 
Eigen 3.3.4 for nodal & & 112 degrees of freedom \\\hline
\textbf{Nodal Total Time} & \textbf{Modal Total Time} & \textbf{Total Time Reduction} \\ \hline
\textbf{1079.63} $\frac{\textrm{seconds}}{\textrm{time-step}}$ & \textbf{67.43} $\frac{\textrm{seconds}}{\textrm{time-step}}$ & $ \sim 16 $ \\ \hline
\textbf{Nodal Vlasov Time} & \textbf{Modal Vlasov Time} & \textbf{Vlasov Time Reduction} \\ \hline
\textbf{1033.89} $\frac{\textrm{seconds}}{\textrm{time-step}}$ & \textbf{60.34} $\frac{\textrm{seconds}}{\textrm{time-step}}$ & $ \sim 17 $ \\ \hline
\end{tabular}
\caption{Summary of the parameters for the numerical experiment to compare the full cost of an alias-free nodal algorithm and an alias-free, quadrature-free, and matrix-free orthonormal, modal algorithm.}
\label{table:VlasovSummaryTable}
\end{center}
\end{table}

Before we conclude this section, the cost of the alias-free nodal scheme is worth additional discussion, most especially whether the alias-free nodal scheme can be optimized to be competitive with the alias-free modal scheme.
The simplest means of further optimization of the nodal algorithm would be vectorize the matrix-vector multiplications over grid elements to produce matrix-matrix products. 
Such an algorithmic rearrangement could lead to a factor of two to four improvement given Eigen benchmarks for matrix-matrix products via more explicitly vectorizable code.
However, with our findings of the significant speed-up from the nodal scheme to the modal scheme, even an aggressive optimization of the nodal scheme derived and implemented in \cite{Juno:2018} would still not be competitive with the modal scheme derived and implemented here.

It is worth further commenting on the sizable amount of research on efficiency improvements to nodal implementations of DG algorithms beyond just an implementation which is amenable to more aggressive vectorization and fast matrix-matrix products.
There exist algorithmic improvements which require the use of tensor product bases, which increases the number of degrees of freedom per cell, but can significantly reduce the number of flops required to perform the quadrature based update using the combination of sum-factorization and a fast Kronecker product\footnote{With tensor product bases and a tensor product quadrature, instead of interpolating a given basis function onto all $N_q = (p+1)^d$ quadrature points in the tensor basis, one can factorize the sums to perform the interpolations and evaluations along the quadrature in a given direction, reducing the number of quadrature evaluations from $\mathcal{O}((p+1)^d)$ to $\mathcal{O}(d (p+1))$.} \cite{Fehn:2019,Kronbichler:2019a,Kronbichler:2019b}.
While these techniques have not yet been applied to higher dimensional kinetic equations, the improvement from $\mathcal{O}((p+1)^{2d})$ operations per cell for the tensor product basis ($N_p = (p+1)^d$) to $\mathcal{O}(d (p+1)^{d+1})$ would be a sizable scaling improvement for five and six dimensional problems. 
However, while it is not necessary to have the same number of quadrature points as basis functions in each direction to exploit the tensor product structure with the combination of sum-factorization and a fast Kronecker product, the optimal scaling quoted still requires this connection between the number of quadrature points and number of basis functions.
Thus, to eliminate aliasing errors, as we require for stable discretization of kinetic equations, the extension to these techniques for when the number of quadrature points is larger than the number of basis functions, and the extra interpolation required, would be a critical component to applying these optimizations to a nodal DG discretization of the kinetic equation.

Such an extension was utilized in \cite{Fehn:2019} to examine the efficiency of sum factorization and fast Kronecker products for over-integration of DG discretizations of compressible and incompressible Navier-Stokes.
While we can only speculate about the resulting algorithm cost of these techniques for an alias-free nodal scheme, we can note the efficiency of our alias-free modal scheme in comparison to the reported efficiency of the compressible Navier-Stokes in \cite{Fehn:2019}.
\cite{Fehn:2019} define the efficiency of their compressible Navier-Stokes solver as
\begin{align}
    E_{op} = \frac{\textrm{\# DOFs}}{\textrm{\# cores} \cdot t_{wall}},
\end{align}
i.e., how many degrees of freedom can be updated per second per core.
Note that this efficiency is defined per scalar value, so although the Navier-Stokes' equations are a vector set of equations, this definition is still amenable to comparisons to a measured efficiency for a scalar partial differential equation such as the Boltzmann (or Vlasov) equation of interest here.
In addition, this efficiency is measured on the complete operator to evaluate the spatial discretization, and thus does not include multiplicative factors from evaluating the spatial discretizations in a multi-stage time-stepping routine, e.g., an explicit Runge-Kutta updater.
In other words, this efficiency comparison is for the cost to take a forward Euler time-step with a DG spatial discretization.
Two caveats to the comparison are that the solver in \cite{Fehn:2019} includes a diffusion operator for the viscosity of the fluid, which is a nontrivial component of the cost, and of course that the Navier-Stokes solver in \cite{Fehn:2019} is only three dimensional.

For the case of the necessary over-integration to resolve quadratic nonlinearities, $N_q = (3p+1)/2$, in each dimension, \cite{Fehn:2019} finds that they can update $10^7$ DOFs per second per core with a $p=2$ tensor product basis in three dimensions.
We can obtain an efficiency from the computational complexity experiment performed in Figure~\ref{fig:algorithmscaling} and we find we can update $\sim 1.67 \times 10^{7}$ DOFs per second per core with the $p=2$ Serendipity basis in five dimensions (2X3V). 
Although this comparison is not one-to-one due to the fundamental differences in the spatial discretizations being compared, \cite{Fehn:2019} includes the cost of a DG discretization of a diffusion operator and we are discretizing a higher dimensional partial differential equation\footnote{Though we note that \gke includes an alias-free modal implementation of a Fokker--Planck operator for collisions in a plasma and involves a DG discretization of a diffusion operator---see \cite{Hakim:2020} for further details. We find that the inclusion of this discrete Fokker--Planck operator roughly doubles the cost of the spatial discretization, and thus the measured efficiency of the complete alias-free DG discretization of the Vlasov--Maxwell--Fokker--Planck system of equations would be $\sim 8 \times 10^6$ DOFs per second per core with p=2 Serendipity elements in five dimensions (2X3V).}, the comparison is nonetheless illustrative that the modal scheme derived and implemented here is competitive with other implementations of DG methods.
And critically, in addition to being competitive cost-wise, we emphasize again that our DG scheme is alias-free, as we require for stable discretizations of kinetic equations.
Having demonstrated the performance of our modal DG algorithm, we turn now to additional implementation details within the \gke~simulation framework.

\section{Gkeyll implementation and parallel scaling}\label{sec:Gkeyll}

The modal kinetic solvers are implemented in \gke, a modern
computational software designed to solve a broad variety of plasma
problems. Though the focus here is ``full'' kinetic equations, 
e.g. the Boltzmann equation or the Vlasov equation, \gke\ also 
contains solvers for reduced kinetic equations such as
gyrokinetics\cite{Hakim:2020ew,Mandell:2020},
as well as for multi-moment multifluid equations\cite{Wang:2020,Dong:2019dz}.

\gke\ uses a number of software innovations which we describe briefly here for completeness. In the context of this
paper, the key features of \gke\ are a low-level infrastructure to
build new solvers and a high-level ``App'' system that
allows putting together solvers to perform a particular class of
simulation. The low-level computational kernels that update a single cell (via
volume and surface DG updates), and compute moments and other quantities
needed in the update sequences, are in C++ and auto-generated using the
Maxima \cite{maxima} CAS. As discussed in the previous section, the use of a CAS allows us to compute most of the
integrals needed in the update analytically, eliminating all
quadrature and unrolling all inner loops to eliminate matrices.

The high-level App system is written in a JIT compiled language,
LuaJIT. Lua is a small, light-weight language that one compiles into the
framework. However, despite its simplicity, LuaJIT is a subtle and powerful language with a
prototype based object system and coroutines that provides great
flexibility in composing complex simulations. Further, the LuaJIT
compiler produces extremely optimized code, often performing at the
level of, or better than, hand-written C,
giving the best of both the worlds: flexibility of a high-level
language as well as speed of a compiled language. We note that \gke\ is less than 8\% (about 36K LOC) hand-written LuaJIT. The rest is autogenerated C++ via the Maxima
CAS. This structure greatly reduces maintenance issues, as one only needs to ensure
that the CAS code is bug-free, rather than coding up all
loops, tensor-tensor products, and quadratures by hand, especially for complex functionality such as the full coupling between the kinetic equation, Maxwell's equations, and a collision operator.

The \gke\ App system greatly simplifies user interaction with the
code. The flexibility of the scripting layer allows the user great
control over the simulation cycle. In fact, every aspect of the
simulation can be controlled by the user without writing any compiled
code, or even the need for a compiler suite. Users can however compile
compute-intensive code by hand and load it into \gke\ using the
LuaJIT FFI. In addition, the App system
streamlines not just the running of a simulation, but also the
manipulation of the data. While post-processing can be done through a
suite of tools called the {\tt postgkyl} package (see \gke\
website\cite{gkeyllgit,gkeylldocs} for details), computationally intensive
analysis techniques can also be run through \gke\ through the App
system.

Input/output is performed with the Adaptable I/O System (ADIOS) \cite{ADIOS:2014} called from LuaJIT.
Using ADIOS, we can both write out data such as velocity moments for analysis with the {\tt postgkyl} package, and also checkpoint/restart simulations for production simulations.
To checkpoint/restart a kinetic simulation, we require the particle distribution function for each of the evolved species and the electromagnetic fields at the last time step. 
The particle distribution functions may be quite large, especially for six dimensional simulations where a modest calculation, $(N_x,N_y, N_z, N_{v_x},N_{v_y},N_{v_z}) = (64,64,64,16,16,16)$ with polynomial order one, $N_p = 64$, and two species, is 1 TB of data.
But the particle distribution function is not just necessary for checkpointing/restarting, it also contains a wealth of data.
Thus, it is the combination of ADIOS and the App system that allows users to read-in output distribution functions, no matter their size, potentially in parallel, and perform computationally intensive but valuable diagnostics such as the field-particle correlation for analyzing the exchange of energy between the plasma and electromagnetic fields \cite{Klein:2016, Klein:2017a, Klein:2017b, Howes:2017}.

An additional software innovation is the two layers of parallel decomposition used by \gke. This multi-layer decomposition is necessary
because there are three grids involved in a kinetic simulation: the
phase-space grid, the configuration-space grid, and the velocity-space
grid. The field solvers work on the configuration-space grid, while
the kinetic equation evolves on the phase-space grid. The coupling
via moments comes from velocity integrals of the distribution function
that lives on the phase-space grid. These grids and various
communication patterns needed to move data between them leads to a
complex use of MPI.

\begin{figure*}
\centering
\includegraphics[width=0.47\textwidth]{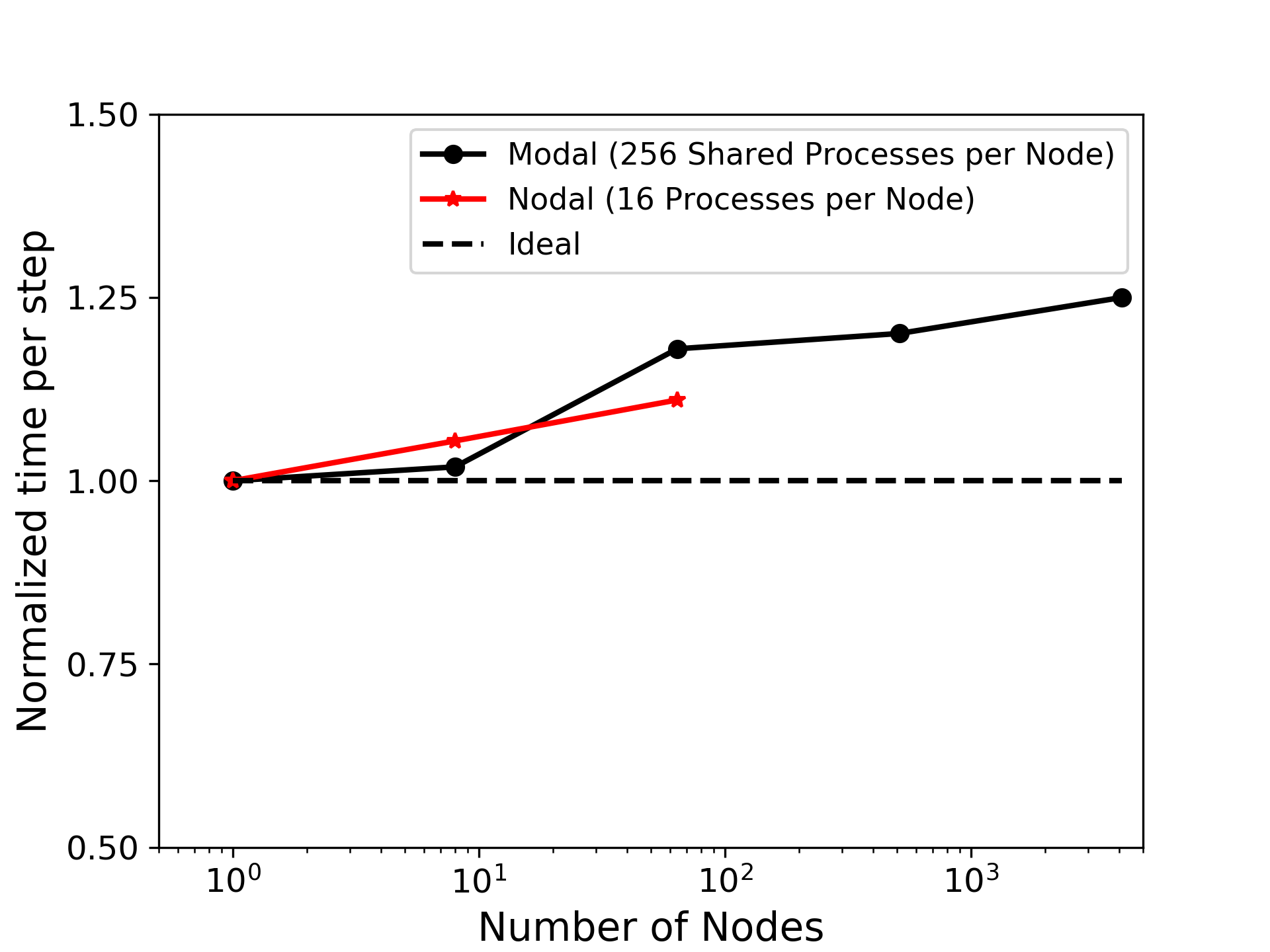}
\includegraphics[width=0.47\textwidth]{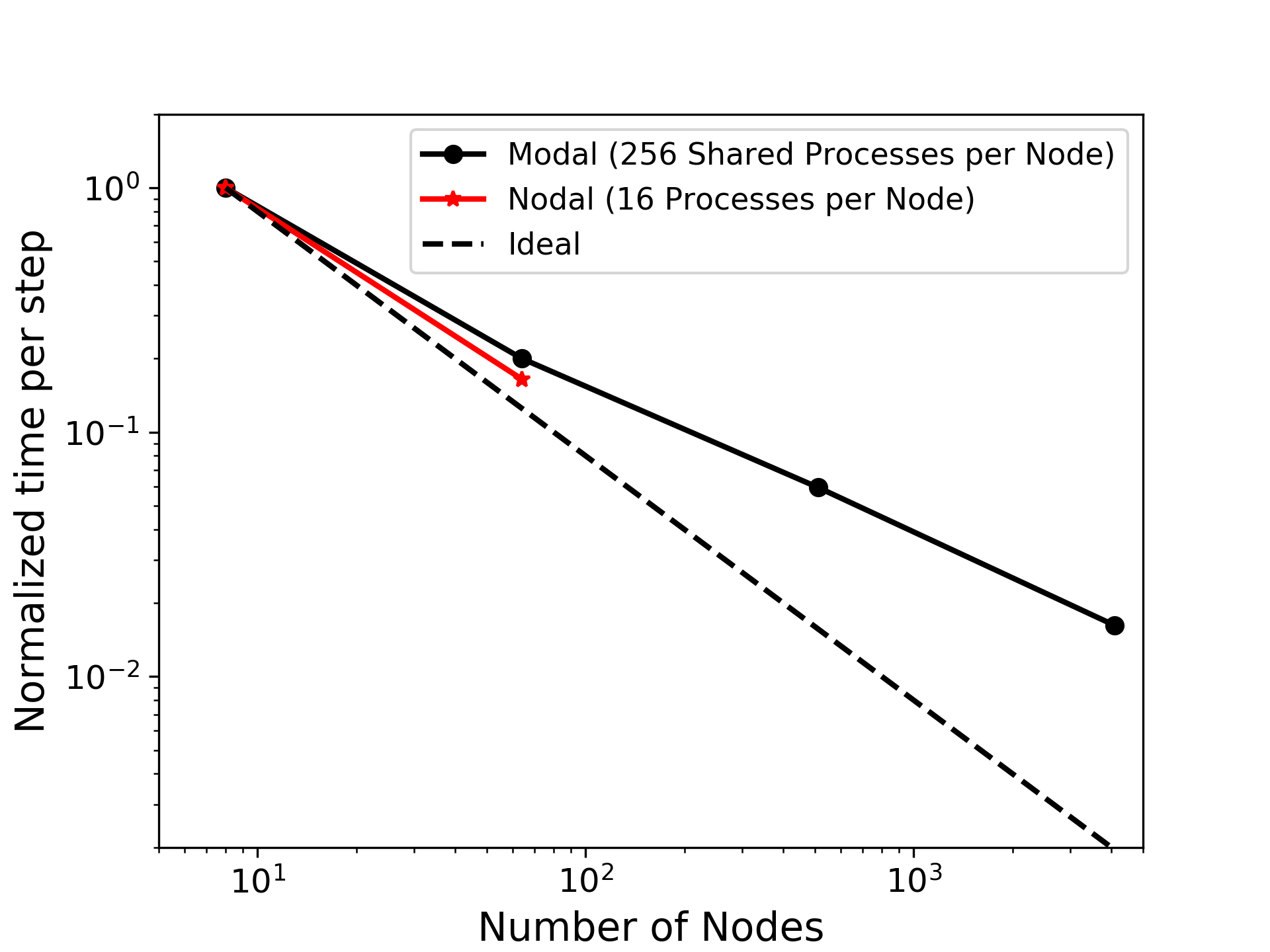}
\caption{Weak (left) and strong (right) scaling results for both the modal (black) and nodal (red) DG algorithms in \gke. To facilitate the scaling performance comparison, we have independently normalized the modal and nodal scalings to their time per step cost on one node for the weak scaling study and eight nodes for the strong scaling study. The modal scaling is performed on Theta at the Argonne Leadership Computing Facility and the nodal scaling is performed on Stampede 1 at the Texas Advanced Computing Center. The scaling of the two algorithms are comparable up to 64 nodes due to the favorable communication pattern of DG algorithms. Importantly though, the modal algorithm has a reduced memory footprint due to the matrix-free implementation of the sparse tensor-tensor multiplies and leverages a larger array of MPI functionality such as MPI-3 shared memory primitives, thus allowing the modal algorithm to more full exploit the many-core parallelism of architectures such as the Knight's Landing Chips on Theta and scale to a larger percentage of the machine.}
\label{fig:gkeyllscaling}
\end{figure*}

\begin{figure}
    \centering
    \includegraphics[width=0.3\textwidth]{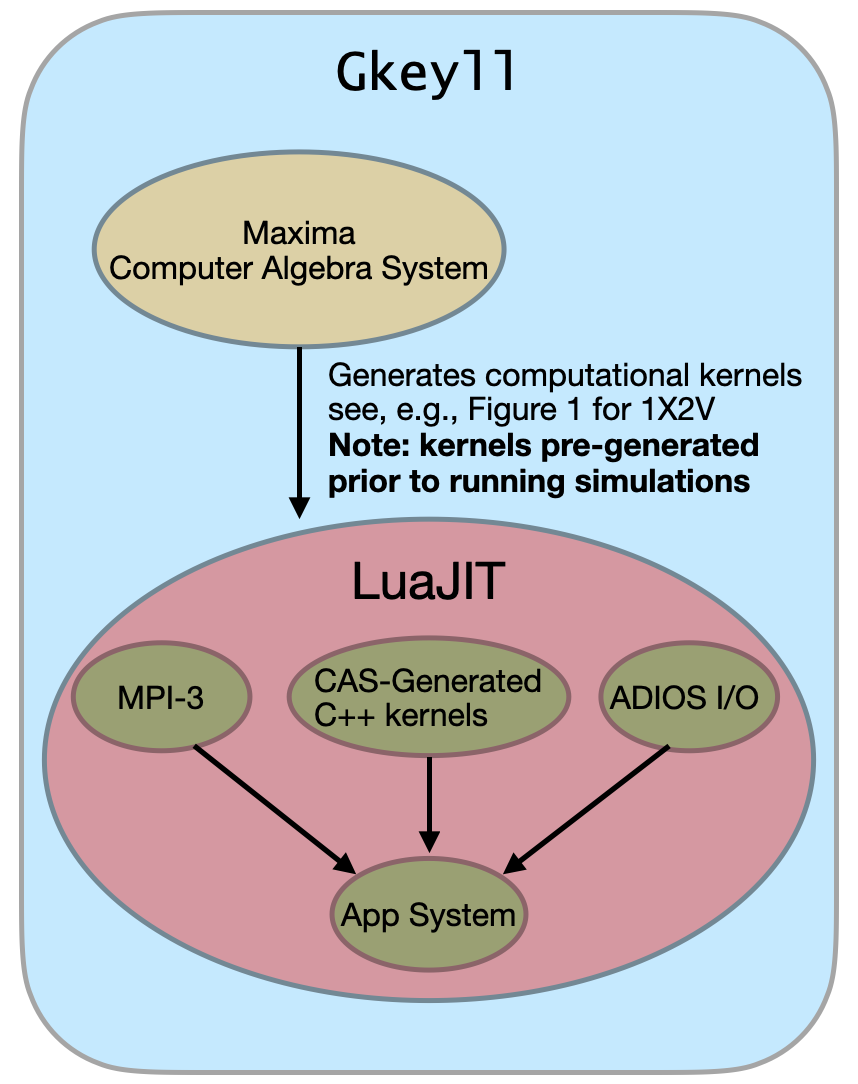}
    \caption{\gke\ program workflow. We note that we use the Maxima computer algebra system to pre-generate the computational kernels, and then wrap the desired functionality in LuaJIT, including what kernels we wish to call based on the desired dimensionality and polynomial order, MPI, and ADIOS for I/O. The App system provides an interface for the user via an additional layer of abstraction for ease of scripting.}
    \label{fig:program-workflow}
\end{figure}

The first level of parallel domain decomposition is in configuration
space. Since the DG algorithm only requires one layer of ghost cells
to compute the surface integrals along each direction, communication
is minimized during the update of the Boltzmann-Maxwell system of
equations. The second level of parallel domain decomposition comes
from a \emph{shared memory} decomposition of the \emph{velocity grid}. For this we use MPI shared-memory primitives to divide the
work in updating a region of velocity space owned by sub-set of the
total number of cores. A further subset of cores on each of these
subsets takes part in the IO and parallel communication.We
use {\tt MPI\_Datatype} objects extensively to avoid unnecessary
copying of data into/out of buffers.
We summarize the \gke\ program workflow in Figure~\ref{fig:program-workflow}.

The advantage of this two-level decomposition is that there is no need
to all-reduce the moment data in velocity space. Further, the use of 
MPI shared-memory primitives eliminates the thread latency common to
thread-based parallelism models such as OpenMP. In other words, our 
parallelism model removes the cost of creating and destroying threads,
while still improving the algorithm's scaling on a single node and    
reducing the amount of memory consumed per node by eliminating the 
need for ghost layers amongst the intra-node work. In fact, due to the high
dimensionality of the problem (say 5D/6D) even a single layer of ghost-cells
need significant memory (as they are 4D/5D) and hence communication time. 
The use of shared-memory on a node significantly reduces memory consumption
(sometimes by $2\times$ or $3\times$).

We note though that in our current shared memory paradigm, we still must select which, and how many, MPI processes will participate in inter-node communication and I/O.
We currently have rank 0 of each collection of shared processes on a node control inter-node communication and I/O, and hence are not guaranteed to saturate the interconnects and maximize bandwidth.
Nonetheless, this shared memory implementation, in addition to other MPI features such as {\tt MPI\_Datatype} further reduces memory consumption, improves our parallel performance, and allows \gke\ to scale on larger clusters.

To quantitatively demonstrate the parallel performance of our pure MPI domain decomposition of the modal DG algorithm, we show in Figure~\ref{fig:gkeyllscaling} the weak and strong scaling of a six dimensional kinetic simulation on the Knight's Landing (KNL) architecture on the Theta supercomputer at the Argonne Leadership Computing Facility.
In addition, we compare the parallel performance of the modal DG algorithm to the scaling of the nodal DG algorithm on a comparably high dimensional problem on Stampede 1 at the Texas Advanced Computing Center.
Note that the parallel performance is tested with the full simulation cycle: Vlasov equation for multiple (two) species, coupled to Maxwell's equations, and using a third order strong-stability preserving Runge-Kutta method, for more than one hundred time-steps to gather good per-time-step statistics.
However, we do not include the cost of I/O using the ADIOS library in this scaling study\footnote{For the largest simulations in the weak scaling study, the particle distribution function is $>$ 4 TB large. While writing out 8 TB of data, 4 TB per distribution function, every one thousand or ten thousand time-steps is an acceptable cost, potentially adding 10-20 percent to the total runtime, writing out this much data has a noticeable impact on the per-time-step performance for this short of a simulation.}.

The initial weak scaling problem for the modal DG algorithm on one node uses $(N_x,N_y, N_z, N_{v_x},N_{v_y},N_{v_z}) =  (8,8,8,16,16,16)$ with polynomial order one, $N_p = 64$.
The problem size is increased by the same amount by doubling the configuration space resolution $N_x, N_y$ and $N_z$, e.g, from $(N_x,N_y,N_z) = (8,8,8)$ to $(N_x,N_y,N_z) = (16,16,16)$ when increasing the number of nodes from one to eight, up to $(N_x,N_y, N_z, N_{v_x},N_{v_y},N_{v_z}) =  (128,128,128,16,16,16)$ on 4096 nodes.
For the nodal DG algorithm, due to its increased expense, we consider a four dimensional, one spatial dimension and three velocity dimensions (1X3V), simulation scaling from $(N_x, N_{v_x},N_{v_y},N_{v_z}) =  (64, 8, 8, 8)$ to $(N_x, N_{v_x},N_{v_y},N_{v_z}) =  (1024, 16, 16, 16)$ with polynomial order four, $N_p = 136$, from one node to 128 nodes.
Note that we are only plotting the performance of the nodal scheme on one node, 8 nodes, and 64 nodes for ease of comparison to the modal scheme.
For the strong scaling problem, we begin with a $(N_x,N_y,N_z,N_{v_x},N_{v_y},N_{v_z}) = (32,32,32,8,8,8)$, polynomial order one, $N_p = 64$, for the modal DG algorithm, and a $(N_x, N_{v_x},N_{v_y},N_{v_z}) =  (256, 16, 16, 16)$, polynomial order four, $N_p = 136$, on eight nodes.
The problem size is then kept fixed and the number of nodes increased, up to 4096 for the modal strong scaling results, and 128 for the nodal strong scaling results.

Up to a modest number of processors, the parallel performance between the nodal and modal scheme are comparable.
This parallel performance is expected for standard DG algorithms, as we require only local data, i.e., one layer of ghost/halo cells, to perform the surface integrals and update the solution.
However, we can scale the modal DG algorithm to a larger percentage of the machine due to both the large array of MPI functionality such as MPI-3 shared memory primitives and {\tt MPI\_Datatype}, and the matrix-free implementation of the computational kernels, which reduces the memory footprint of the modal DG algorithm.
While the nodal DG algorithm does not have the same spectrum of implemented MPI features, we expect the modal DG algorithm would still exceed the nodal DG algorithm parallel performance because of this reduced memory footprint, as the nodal DG algorithm would require the sharing of modestly large matrix data structures that encode the necessary high-order quadrature to eliminate aliasing errors.
Thus, the sharing of these dense data structures would not be guaranteed to improve parallel performance because caching and efficient memory movement is still a significant challenge with the alias-free nodal algorithm.

The shared memory implementation of the modal DG algorithm has additional strengths.
We note that even though we are not using a thread-based model for parallel programming on a node, we can still take advantage of all 256 ``threads'' on the KNL chip by specifying at runtime that we are using 256 shared MPI processes
These options provide significant benefit for our DG algorithm, as the use of all 256 ``threads'' not only allows us to divide the work amongst a larger number of processes, using multiple ``threads'' per core exposes a much greater degree of instruction level parallelism, reducing the cycles per instruction and leading to greater floating point efficiency.
Instruction level parallelism is particularly useful for our application, as the instructions of our unrolled sparse tensor-tensor products such as the computational kernel shown in Figure~\ref{fig:1X2V-p1-vol-update}, while sparse relative to the nodal algorithm, are still dense instruction sets.
Thus, while we have reduced the memory footprint significantly with our modal DG implementation compared to the nodal DG implementation, we still must take care to avoid wasting clock cycles fetching instructions to complete the sparse tensor-tensor products by maximizing instruction level parallelism.

In fact, instruction level parallelism is the reason our weak scaling is more favorable than our strong scaling, as our weak scaling maintains enough work per node to extract a larger efficiency using 256 shared MPI processes; whereas, the fixed problem size is not enough work on a decomposition using the full machine. 
Thus, there is degradation of performance within the node in the strong scaling case, even though communication is minimized by our DG algorithm.
In the weak scaling study, we find at worst 25 percent of the per-time-step cost is spent in halo/ghost cell exchange for a problem a factor of four thousand times larger (a factor of sixteen increased resolution in each configuration space dimension), while we find the ratio of communication time to computation time per time step for the strong scaling time to be larger.
We expect the simulation in the strong scaling study to be a factor of five hundred times faster per time step, but instead only find a factor of sixty times speed-up. 
At each factor of eight increase in the number of nodes, we gain only a factor of four speed up, so the communication time to computation time ratio is compounding by an additional fifty percent in each increment of the node count.
The on-node performance degradation from less instruction-level parallelism, combined with a larger amount of data transfer as the ratio of ghost cells/interior cells increases, leads to upwards of 80 percent of the per-time-step cost spent communicating halo/ghost cells on the 4096 node strong scaling simulation, though the run time is still reduced by a factor of sixty from the eight node strong scaling simulation.

Due to the memory requirements of solving a six dimensional PDE, we are limited on the base problem size we can choose for strong scaling.
In fact, the memory requirements for solving a six dimensional PDE ultimately make weak scaling of principal interest, as access to a larger amount of memory on distributed memory clusters is what allows for the calculation of interesting kinetic systems.
We note though that we are actively pursuing improvements to the demonstrated scalings by more optimized data transfers using multiple MPI processes per node in communication and MPI neighborhood collectives, such as {\tt MPI\_Ineighbor\_alltoall}, and by overlapping communication and computations.
These improvements should lead to near-ideal weak scaling and significantly less performance degradation in strong scaling studies.
Nevertheless, the modal DG algorithm, using a large suite of MPI functionality including the MPI-3 shared-memory primitives and {\tt MPI\_Datatype}, has good scaling up to the machine size (4096 KNL nodes and $>$1 million MPI processes).

\begin{figure*}
    \centering
    \includegraphics[width=0.8\textwidth]{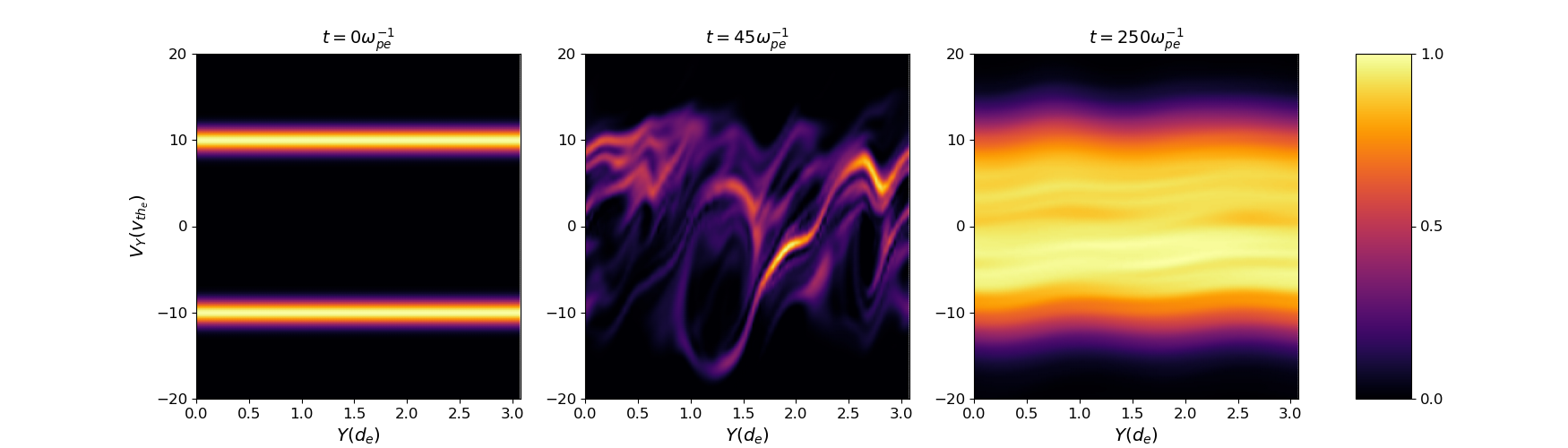}
    \includegraphics[width=0.8\textwidth]{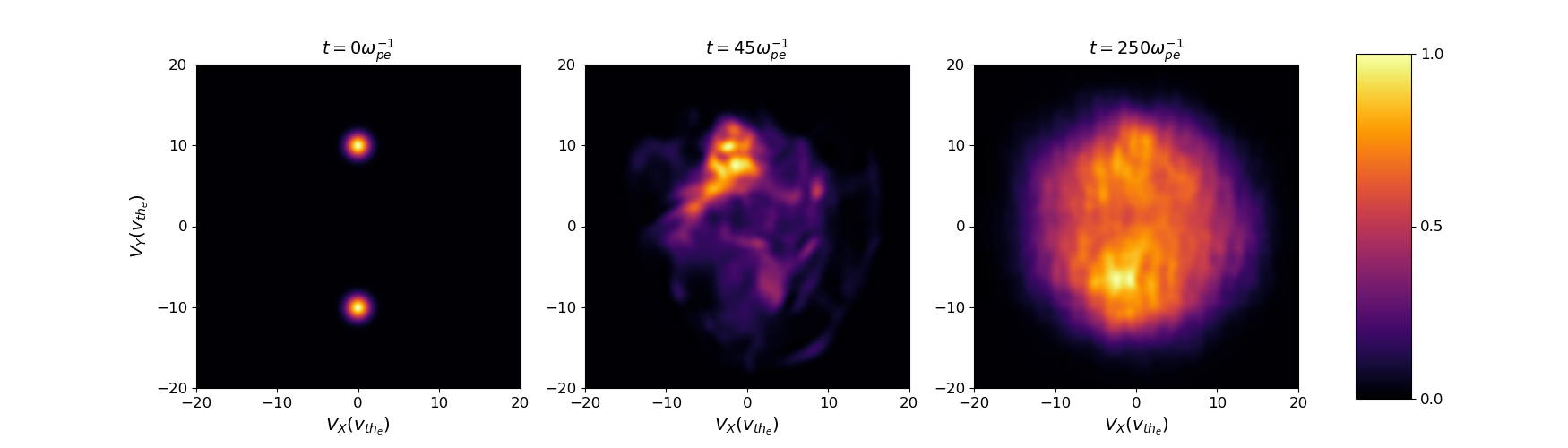}
    \caption{Evolution of an unstable plasma system driven by counter-streaming beams of electrons. We show cuts of the electrons in $y-v_y$ (top) and $v_x-v_y$ (bottom), at three different times, the initial condition, the beginning of nonlinear saturation, and the end of the simulation. These distribution function slices demonstrate both the velocity space structure generated by the kinetic instabilities present, as well as the utility of a continuum kinetic method in representing this velocity space structure. See\cite{Skoutnev:2019,Juno:2020} for details.}
    \label{fig:distcuts}
\end{figure*}

\section{Example Simulations}

We now briefly show the results of an example simulation run with the modal DG algorithm presented in this paper.
We repeat the calculation of previous publications which used the modal DG algorithm in \gke \cite{Skoutnev:2019,Juno:2020}.
This particular simulation demonstrates the utility of a continuum kinetic approach, as the high fidelity representation of the particle distribution function provides critical insights for our understanding of the dynamics of this kinetic system, in this case a collisionless plasma.
In fact, the authors of \cite{Juno:2020} demonstrated in comparing the results of \cite{Skoutnev:2019} to a particle-based method found that the noise inherent to the PIC algorithm can pollute the results of the simulations.
Additionally, while similar simulations were performed with the nodal algorithm in \cite{Juno:2018} using \gke\ in \cite{Cagas:2017a,Cagas:2017b}, the modal algorithm employed here and in \cite{Skoutnev:2019,Juno:2020} obtains results an order of magnitude faster.

The setup is an electron-proton plasma in two spatial dimensions, two velocity dimensions (2X2V), with the electron population initially divided amongst two counter-streaming beams.
These counter-streaming beams serve as a source of free-energy for a zoo of plasma instabilities, including two-stream, filamentation, and hybrid two-stream-filamentation modes \cite{Bret:2009}.
In the limits explored in \cite{Skoutnev:2019}, the authors found that as the beam velocity became both more nonrelativistic and colder, such that the beam's initial energy was dominantly kinetic energy, a large spectrum of hybrid two-stream-filamentation, or oblique, modes all had comparable growth rates.
With multiple unstable modes all growing and vying for dominance, the nonlinear saturation of these instabilities led to a highly dynamic phase space.
This highly dynamic phase space had a significant impact on the late-time evolution of the plasma, with collisionless damping of the saturated modes depleting the generated electromagnetic energy of the unstable modes, and leading to overall energy conversion from kinetic to electromagnetic to thermal due to the instability dynamics.

We show in Figure~\ref{fig:distcuts} the electron distribution function at three different times, the initial condition, the time of nonlinear saturation when the electromagnetic energy peaks, and the end of the simulation, with two different slices of phase space, $y-v_y$ (top) and $v_x-v_y$ (bottom).
These distribution function slices demonstrate the phase space structure that can be represented with a continuum kinetic method such as the modal DG algorithm presented here.
We emphasize that this phase space structure is an important component of the dynamics, and said phase space structure was leveraged in \cite{Skoutnev:2019} to determine why the macroscopic energetics, the partition between electromagnetic and electron energy, differed from previous simulations of these types of instabilities \cite{Kato:2008}.
In this regard, the fact that particle noise can affect the simulation dynamics by reducing the effective phase space resolution of the simulation and polluting the late time evolution, as demonstrated by \cite{Juno:2020} when comparing to the results of \cite{Skoutnev:2019} using a PIC method, leads us to highlight the utility of the modal DG algorithm presented in this paper.
Not only is the modal DG algorithm presented here an order of magnitude faster than the previous nodal algorithm described in \cite{Juno:2018}, while maintaining the same level of accuracy and eliminating aliasing errors, a continuum kinetic method provides additional fundamental insights into the evolution of kinetic systems such as collisionless plasmas by both allowing us to directly diagnose the physics using the distribution function dynamics and eliminating the counting noise which can plague particle-based methods such as the PIC algorithm.

\section{Conclusion}

In this paper we have presented, to our knowledge, the first alias-free, matrix-free and quadrature-free scheme for continuum simulation of kinetic problems. Kinetic problems are characterised by delicate field-particle energy exchange that requires great care to ensure that aliasing errors do not modify the physics contained in the system. Further, as kinetic systems evolve in high dimension phase-space (5D/6D) it is important to ensure that the computational cost is minimized, while still retaining accuracy and convergence order. Our modal DG scheme achieves this by computing all needed volume and surface integrals analytically using a computer algebra system and generating the computational kernels automatically. These kernels leverage the sparsity of the tensor-tensor convolutions with a modal, orthonormal basis, unroll all loops and eliminate the need for matrices, and consolidate common expressions with common factors ``pulled out''. This leads to dramatic reduction in FLOPs and data movement, significantly speeding up computing time compared to a nodal DG code even when the latter uses highly optimized linear algebra libraries. Critically, despite the analytical elimination of aliasing, we still obtain sub-quadratic scaling of cost with degrees-of-freedom per cell.

Our scheme is implemented in the flexible, open-source computational plasma physics framework, \gke. This framework allows flexible construction of simulations using a powerful ``App'' system. \gke\ is mostly written in LuaJIT, a JIT compiled language, with key computational kernels written in C++, pre-generated by the Maxima computer algebra system. A hybrid MPI-shared-MPI domain decomposition allows us to reduce communication within nodes and ensures almost linear scaling on a single node, while retaining excellent scaling properties across nodes. We have demonstrated this on the Theta supercomputer all the way up to the full machine ($>$~1 million MPI processes).

Our present algorithmic work is focused on two areas: adding a multi-moment model coupling to the kinetics that will lead to a unique hybrid moment-kinetic simulation capability (most hybrid PIC codes assume massless, isothermal electrons), and a novel recovery based DG scheme that will further increase accuracy \cite{VanLeer:2005,VanLeer:2007}, reducing resolution requirements. The recovery based approach is very promising as it may allow achieving, for example, $4th$ order convergence with just $p=1$ DG basis functions where traditional DG schemes obtain $p+1$ order convergence for $p$-th order basis. Such increase in accuracy can allow use of coarser meshes, further dramatically reducing the computation cost for 5D/6D problems. And with the use of the Maxima CAS to analytically evaluate the recovery polynomials and auto-generate the computational kernels, the added complexity of the recovery-based approach in higher dimensions can be significantly mitigated. Combined with the flexibility of the \gke\ code these innovations will enable larger problems of interest in a broad array of fields.

\section*{Acknowledgment}
This work used the Extreme Science and Engineering Discovery Environment (XSEDE), which is supported by National Science Foundation grant number ACI-1548562 and resources of the Argonne Leadership Computing Facility, which is a DOE Office of Science User Facility supported under Contract DE-AC02-06CH11357. We thank the \gke\ team for contributions to various parts of the code and the work we have presented here. In particular, we thank Mana Francisquez for help in implementing the collision operators, moment computations and other core code, Noah Mandell for help in high-level App system (and authoring the gyrokinetic solvers) and Petr Cagas for implementing the post-processing tools (and BGK collision operators and boundary conditions). We also thank Jason TenBarge, Greg Hammett and Bill Dorland for extensive discussion on various aspects of the physics of the Vlasov-Maxwell system.



\bibliographystyle{plain}
\bibliography{abbrev,hakim-juno-sc}

\end{document}